\documentclass[prl,twocolumn,superscriptaddress]{revtex4-2}

\usepackage[utf8]{inputenc}
\usepackage{graphicx}
\usepackage{hyperref}
\usepackage{amsmath}
\usepackage{amssymb}
\usepackage{tabularx}
\usepackage{booktabs}
\usepackage{lineno}
\usepackage{xcolor}
\usepackage{lineno}
\usepackage{bm}

\hypersetup{
	colorlinks=true,
	linkcolor=blue,
	filecolor=blue,
	urlcolor=blue,
	citecolor=blue,
}

\begin{document}
\title{5d orbital Induced Room Temperature Quantum Anomalous Hall Effect in TbCl}

\author{Jianqi Zhong$^{\#}$}
\address{Wuhan National High Magnetic Field Center $\&$ School of Physics, Huazhong University of Science and Technology, Wuhan 430074, China}

\author{Jianzhou Zhao$^{\#}$}
\address{Co-Innovation Center for New Energetic Materials, Southwest University of Science and Technology, Mianyang 621010, China}

\author{Jinyu Zou$^{*}$}
\address{Wuhan National High Magnetic Field Center $\&$ School of Physics, Huazhong University of Science and Technology, Wuhan 430074, China}

\author{Gang Xu$^{*}$}
\address{Wuhan National High Magnetic Field Center $\&$ School of Physics, Huazhong University of Science and Technology, Wuhan 430074, China}
\address{Institute for Quantum Science and Engineering, Huazhong University of Science and Technology, Wuhan, 430074, China}
\address{Wuhan Institute of Quantum Technology, Wuhan, 430074, China \\\ \text{Email:jyzou@hust.edu.cn} \\\ \text{Email:gangxu@hust.edu.cn} \\\ \text{\#:These authors made equal contributions to this work.}}

\begin{abstract}

Following the experimental realization of Quantum anomalous Hall (QAH) effect in thin films of chromium-doped (Bi,Sb)$_2$Te$_3$, enhancing the work temperature of QAH effect has emerged as a significant and challenging task. Here we demonstrate monolayer TbCl as a promising candidate to realize the room temperature QAH effect. Using DFT+U method, double checked by HSE06 and DMFT calculations, we identify the Hall conductivity $G = -e^2/h$ per layer in three-dimensional ferromagnetic insulator TbCl, which is a weakly stacking of QAH layers. The monolayer TbCl inherits the magnetic and topological properties, exhibiting the QAH effect with Chern number $C$=-1. The large topological band gap reaches 42.8 meV, which is beyond room temperatue. The extended 5$d$ electrons lead to sizable exchange and superexchange interactions, resulting in a high Curie temperature $T_c$$\sim$457K. All these features demonstrate that monolayer TbCl will provide an ideal platform to realize the room temperature QAH effect.

\end{abstract}
\maketitle

\section{Introduction}
Quantum anomalous Hall (QAH) effect is the most profound topological quantum state and triggers the flourish of topological matters since proposed by Haldane~\cite{haldane1988model, 10.1093/nsr/nwt029,weng2015quantum,liu2016quantum,chang2023colloquium, hasan2010colloquium,qi2011topological,tokura2019magnetic,wang2017topological}. Its quantized Hall conductivity $\sigma_H=C\frac{e^2}{h}$ is characterized by a finite Chern number $C$ of the band structure, corresponding to the non-dissipative chiral edge states which have great prospective applications in low-power-consumption electronics. However, the accomplishment of QAH effect so far requires extremely low temperature, i.e. $\backsim$30 mK in magnetically doped topological insulators (TI)~\cite{chang2013experimental,chang2015high,mogi2015magnetic,ou2018enhancing}, $\backsim$1.4 K in the intrinsic magnetic TI MnBi$_2$Te$_4$~\cite{deng2020quantum}, $\backsim$2.5 K in the moiré systems~\cite{serlin2020intrinsic,li2021quantum}, thus hinders the application in reality. Enhancing the working temperature is the key step for the future advancement of this field, which encounters challenges primarily from two fronts. The first challenge is to enhance the Curie temperatures ($T_c$) of magnetic materials, while the second is to enhance the topological gap of the band structure. 
The pursuing of room temperature QAH effect encourages many theoretical efforts on the ferromagnetic candidates with high $T_c$ and sizable topological gaps~\cite{wang2018high,you2019two,sun2020intrinsic,kong2018quantum,li2020monolayer,li2020high,jiang2018screening,chen2021high, li2024weyl,xu2015intrinsic,Xu2024}. 
To date, an experimentally accessible candidate for room temperature QAH effect is still in sought.

Recent studies on topological classification ~\cite{turner2012quantized,ono2018unified,peng2022topological,nie2019topological} suggest that the time-reversal symmetry breaking three-dimensional (3D) insulators could exhibit the weak topological properties, featuring integer-quantized Hall conductivity per layer. This topological phase can be understood as a stacking of QAH layers by weak interlayer interaction, and it has been referred to as a ``3D weak Chern insulator" ~\cite{xu2015,xu2011}. When fabricated into thin film, its Hall conductivity is quantized and proportional to the number of layers. Such feature suggests a promising route to realize the QAH effect by exfoliating the 3D weak Chern insulator to obtain the QAH layer.

In this work, we predict the Lanthanide halide TbCl as such a prominent candidate based on first-principles calculations. The van der Waals (vdW) layered bulk TbCl is identified by our calculation as a ferromagnetic insulator with out-of-plane magnetic moments, which exhibits the 3D weak Chern insulator phase characterized by the Hall conductivity $G=-e^2/h$ per layer. Due to the small exfoliation energy about 0.24 J/m$^2$, the bulk TbCl can be exfoliated easily into monolayer and the monolayer is confirmed to be dynamically stable by the phonon spectrum. Notably, the monolayer inherits the ferromagnetism (FM) of the bulk and exhibits the high $T_c\thickapprox$ 457 K because of the sizable exchange and superexchange interactions originated from the extended 5$d$ electrons. The band structure calculation identifies the monolayer TbCl as QAH insulator featured by Chern number $C$= -1 with large gap 42.8 meV, which is obviously beyond the room temperature. These results strongly demonstrate that the monolayer TbCl is an ideal candidate for achieving the room temperature QAH effect, due to its small exfoliation energy, high $T_c$, and large topological gap.

\section{Results}
\subsection{Electronic and topological properties of bulk TbCl}
\begin{figure}[t]
	\centering
	\includegraphics[width=0.48\textwidth]{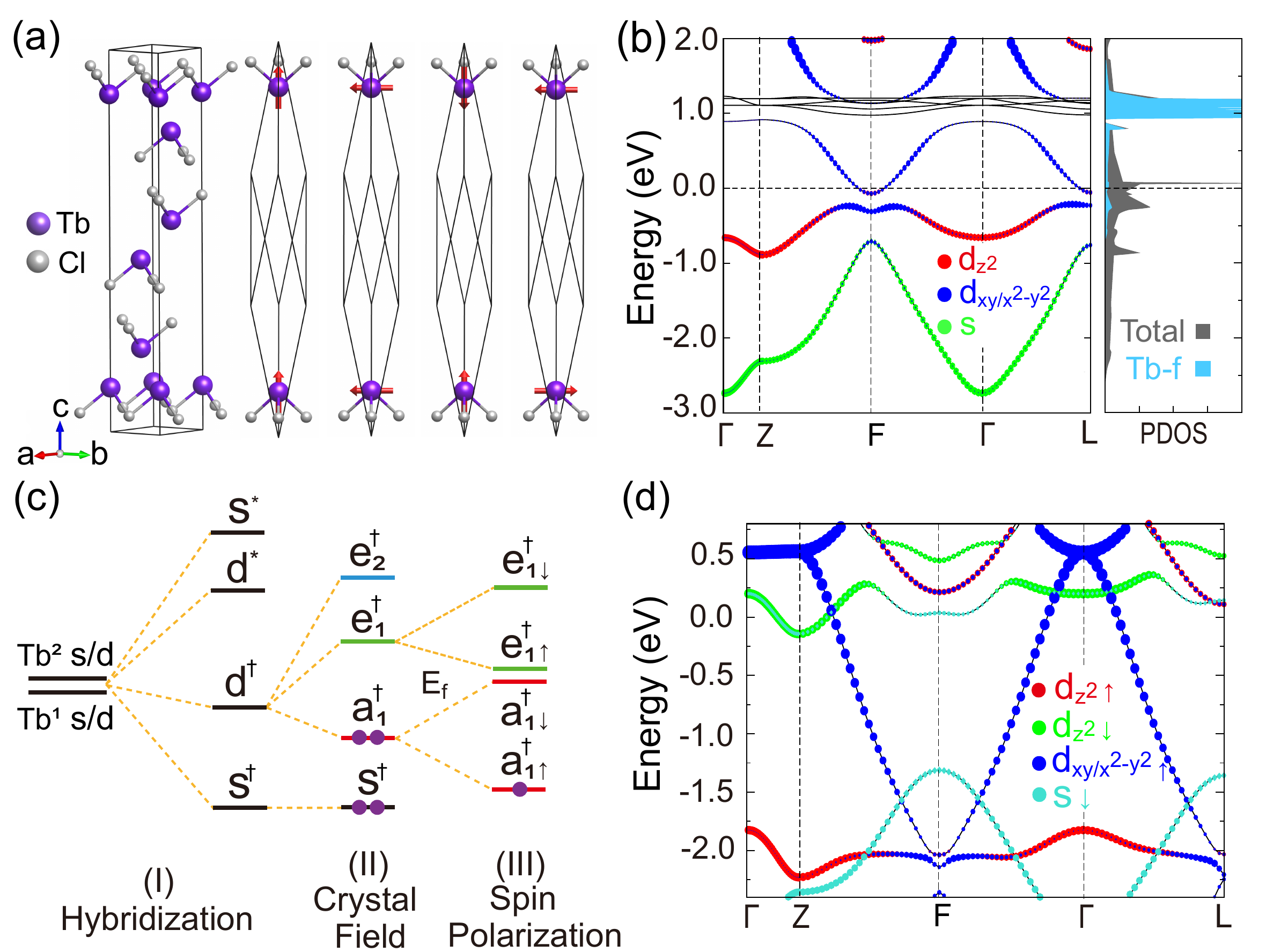}
	\caption{Structure and orbital analysis of TbCl. (a) Crystal structure of the bulk TbCl and its magnetic structures in the primitive cell. (b) HSE06 calculated  DOS (Right panel) and band structure (Left panel) for the bulk in the nomagnetic state. Red, blue and green represent the projection on  $d_{z^2}$, $d_{xy/x^2-y^2}$ and $s$ orbitals of Tb. (c) Splitting of the 5$d$ and 6$s$ orbitals due to the hybridization, crystal field and spin polarization. (d) The HSE06 calculated FM band structure without SOC.
	}
	\label{fig1}
\end{figure}

The experiment has synthesized the bulk TbCl~\cite{simon1976monochlorides}, which presents an ABC-type stacking pattern along the $c$ direction with the $R \bar{3}m$ ($D_{3d}$) symmetry, as shown in Fig.~\ref{fig1}(a). The lattice constants are determined as $a$=3.787 {\AA} and $c$=27.461 {\AA}. Tb and Cl both sit at Wyckoff positions 6$c$ with coordinates (0,0,0.116) and (0,0,0.385).

We perform the HSE06 calculation to show the band and density of states (DOS) for the bulk nomagnetic state in Fig.~\ref{fig1}(b). 
The $4f$ orbitals are pushed away from the Fermi level by the strong Coulomb interaction, and eight $f$ orbitals on each Tb are occupied. Thus the Fermi level is dominated by the $d_{z^2}$, $d_{xy/x^2-y^2}$ and $s$ orbitals. These results are further confirmed by the DMFT calculation as shown in Supplementary Figure 1 in the supplementary~\cite{Zhongsupplemental}. To better understand the 5$d$ and 6$s$ orbital bands, we further study the evolution of the orbitals at the $\Gamma$ point as shown in Fig. \ref{fig1}(c). The neighboring Tb atoms, due to the short distance (3.54 {\AA}), hybridize with each other to form the lower bonding states $s^\dagger/d^\dagger$ and higher antibonding states $s^\ast/d^\ast$ (Stage I). Among these, the bonding orbital $s^\dagger$ is fully occupied leading to the $6s^1$/Tb and the bonding orbitals $d^\dagger$ near the Fermi level are further split by the trigonal crystal filed into the occupied singlet $a^\dagger_1$ ($d^\dagger_{z^2}$), fully empty doublets $e^\dagger_1$ ($d^\dagger_{xy/x^2-y^2 }$) and $e^\dagger_2$ ($d^\dagger_{xz/yz}$) (Stage II). 
The resulting 4$f^8$5$d^1$6$s^1$ configuration agrees with the previous observation of Tb$^{+1}$ ion very well~\cite{li2021monovalent}.
Subsequently, the spin splitting of 5$d$ orbitals (Stage III) further pushes the $a^\dagger_1{\uparrow}$ branch away from the Fermi level to make it fully occupied, while moves $a^\dagger_1 {\downarrow}$ and $e^\dagger_1 {\uparrow}$ branch toward the Fermi level and cross with each other as shown in Fig. \ref{fig1} (d), which is the  HSE06 calculated band structure of FM configuration without spin-orbital coupling (SOC). 
We employ both HSE06 and DFT+U methods to capture the correlation effects of 5d and 4f electrons. As shown in Fig. S2~\cite{Zhongsupplemental}, DFT+U with U=6 eV yields electronic structures in good agreement with HSE06. Therefore, we adopt U=6 eV for all subsequent DFT+U calculations unless otherwise noted.

Table \ref{tab1} lists the calculated difference energy of four different magnetic structures (Fig.\ref{fig1}(a)), among which FM-z is the ground state with 4.2 meV$/f.u.$ lower energy due to magnetic anisotropy. The spin magnetic moment is found to be $\approx$6.36 $\mu_B$/Tb, which is understood as follows. Firstly, the eight 4$f$ electrons on each Tb occupy $\uparrow^7\downarrow^1$ due to Hund’s coupling, and contribute 6 $\mu_B$/Tb. 
Secondly, the partial occupation of $d^\dagger_{z^2}\downarrow$ reduces the contribution of the spin-up  5$d$ orbitals. 
Thus, the totall spin magnetic moment is finally reduced to be about 6.36 $\mu_B$/Tb.

\begin{figure}[t]
	\centering
	\includegraphics[width=0.48\textwidth]{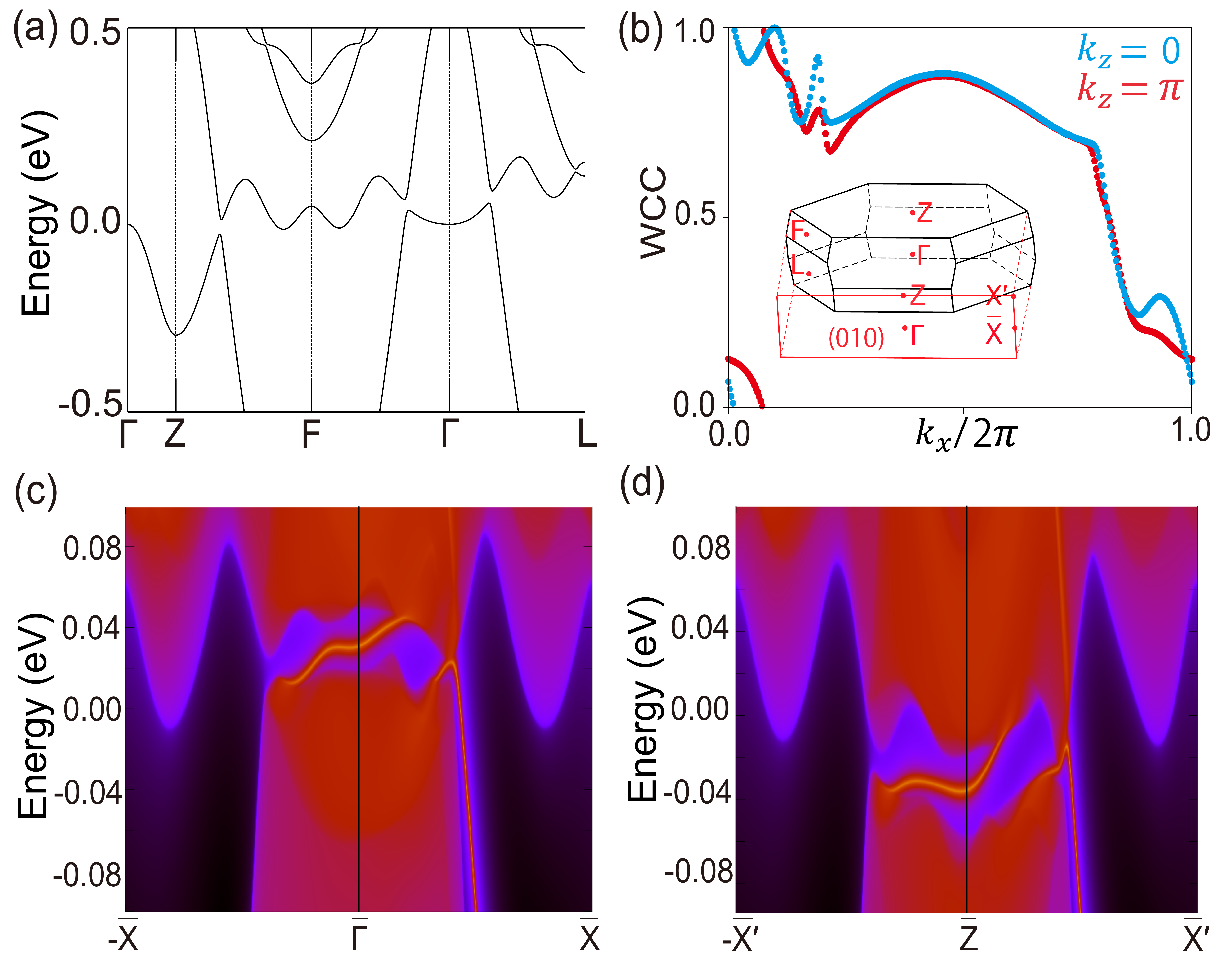}
	\caption{Band structure and topological properties of bulk TbCl. (a) is the DFT+U band structure with SOC in FM-z state at $U=6$ eV. (b) The evolution of Wannier charge center (WCC) along $k_x$ at $k_z=0$ (blue dots) and $k_z=\pi$ (red dots) plane, respectively. The inset is the Brillouin zone and the projected surface Brillouin zones of (010) plane. (c) and (d) are the surface state on (010) plane at $k_z=0$ and $k_z=\pi$, respectively.
	}
	\label{fig2}
\end{figure}

We further discuss the topological properties of the bulk TbCl. Fig. \ref{fig1} (d) shows the band inversion between $d^\dagger_{z^2}\downarrow$ and $d^\dagger_{xy/x^2-y^2}\uparrow$ around F point. The SOC can open a topologically nontrivial gap, as shown in Fig. \ref{fig2}(a). 
We calculate the evolution of the Wannier charge centers to abtain the Chern number in $k_z=0$ and $k_z=\pi$ planes, which are found to be both $C=-1$ as demonstrated in Fig. \ref{fig2}(b). Such results mean that the bulk TbCl is a 3D weak Chern insulator stacked by the QAH layers with Hall conductivity $G=-e^2/h$ per layer, which can be also confirmed by the chiral edge state in the (010) surface along $\bar{\Gamma}-\bar{X}$ and $\bar{Z}-\bar{X'}$ as shown in Fig. \ref{fig2}(c) and (d) respectively.

\subsection{Room temperature QAH effect in monolayer TbCl}
\begin{figure}[t]
	\centering
	\includegraphics[width=0.48\textwidth]{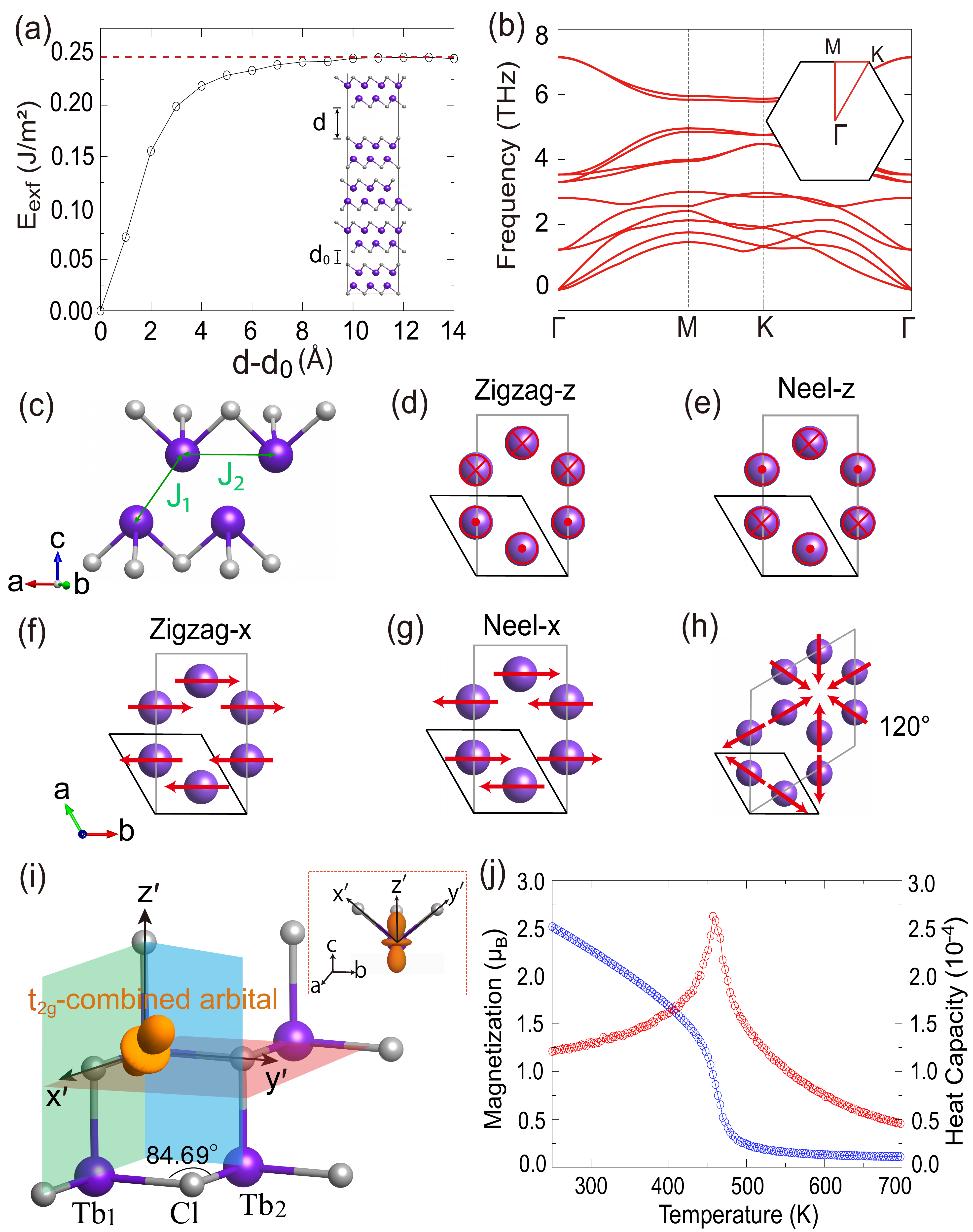}
	\caption{The stability and magnetism of the TbCl monolayer. (a) Calculated exfoliation energy $vs$. separation distance $(d-d_{0})$. The inset shows the side view of the five-layer slab model. (b) Phonon spectra of the TbCl monolayer. The inset shows the Brillouin zone. (c) Side view of monolayer TbCl. $J_1$ and $J_2$ are the nearest and next nearest neighboring exchange coupling, respectively. (d)-(h) The magnetic configuration of Zigzag-z, Neel-z, Zigzag-x, Neel-x and 120$^\circ$ type antiferromagnetic structure. (i) Illustration of the local coordinates $x’-y’-z’$ along the Tb-Cl bonds. The light red, blue and green colors label the $x^{\prime} y^{\prime}$, $x^{\prime} z^{\prime}$, $y^{\prime} z^{\prime}$ planes, respectively. The global $d_{z^2}$ orbital is represented by the orange ball. The inset shows the side view of the local coordinates. (j) The magnetization (blue dot-line) and specific heat (red dot-line) depending on the temperature. 
	}
	\label{fig3}
\end{figure}

The topological properties of the 3D vdW bulk TbCl with the Hall conductivity $G=-e^2/h$ per layer of the crystal suggest it as a perfect candidate to fabricate the 2D QAH system\cite{xu2015intrinsic,turner2012quantized,ono2018unified,peng2022topological}. The exfoliation energy of monolayer TbCl, determined using a five-layer slab model as shown in Fig.  \ref{fig3}(a), is 0.24 J/m$^2$, which is noticeably lower than the experimental value for graphite ($\backsim$0.32 J/m$^2$) and H-MoS$_2$ ($\backsim$0.29 J/m$^2$)~\cite{zacharia2004interlayer,jung2018rigorous}. Such a low exfoliation energy suggests that monolayer TbCl can be more easily exfoliated from its bulk counterpart, indicating its excellent experimental accessibility.
The monolayer TbCl belongs to space group $P \bar{3}m1$ (no.164), where the Tb and Cl atoms are located at 2$c$ and 2$a$ Wyckoff position respectively, as show in Fig. \ref{fig3}(c).
The calculated phonon spectrum demonstrates that the monolayer TbCl is dynamically stable as shown in Fig. \ref{fig3}(b). Additionally, the formation energy of monolayer TbCl is calculated to be -2.8908 eV/atom, and ab initio molecular dynamics (AIMD) simulations further confirm its thermal stability at both 300 K and 600 K\cite{Zhongsupplemental}.

To determine whether the bulk magnetism is inherited by the monolayer, we compare the energies of several representative magnetic structures including the FM, Neel, Zigzag and the 120$^\circ$-AFM, as illustrated in Fig. \ref{fig3} (d-h).
The magnetic anisotropic energy 3.8 meV$/f.u.$ is obtained (see Table. S1 in SM), consistent with the bulk magnetic anisotropic energy. Thus we only list the z-oriented configurations and 120$^\circ$-AFM in Table \ref{tab2}, which identifies FM-z as the ground state, consistent with the bulk magnetism.
Based on these results, we estimate the magnetic transition temperature by employing the Heisenberg model up to next nearest exchange coupling: 
\begin{equation}\label{Hmodel}
  H=-J_1\sum_{\langle i,j\rangle}\bm S_i\cdot \bm S_j-J_2\sum_{\langle\langle i,j\rangle\rangle}\bm S_i\cdot \bm S_j-D\sum_i(S_i^z)^2.\tag{3}
\end{equation}
where $J_1$ and $J_2$ are the nearest (interlayer) and next nearest neighboring (intralayer) exchange coupling as labeled in Fig. \ref{fig3}(c), $\langle i,j\rangle$ and $\langle\langle i,j\rangle\rangle$ represent the summation over nearest and next nearest neighbors. $\bm S$ ($|S|$=3.5) is the spin angular momentum of Tb atom and $S_i^z$ is the z component. $D$ represents the anisotropy energy. 
Using the energy of different configurations, we can immediately extract the values of the $J_1$, $J_2$ and $D$ as listed in Table \ref{tab2}.
With these parameters, the Curie temperature is evaluated by Monte Carlo simulation as $T_c\thickapprox$ 457 K (see details in the SM~\cite{Zhongsupplemental}), which is obtained by plotting the temperature dependent magnetization and specific heat in Fig. \ref{fig3}(j). 
Such significant high $T_c$, even exceeding room temperature, is because that both $J_1$ and $J_2$ are FM and sizable that are originated from the more extended 5$d$ orbitals as analyzed in the next paragraph, which usually leads to an order of magnitude larger exchange coupling comparing to that of the 4$f$ orbital-dominated rare earth magnets like GdI$_3$~\cite{you2021peierls} and  EuS~\cite{moruzzi1963specific,guntherodt1976configurations}, and gives rise to transition temperatures in the hundreds of Kelvin, as reported in LaCl ($\sim$260K)~\cite{jiang2018screening} and GdI$_2$ ($\sim$241K)~\cite{Felser1999,xu2015,wang2020prediction}.

The FM mechanism of $J_1$ and $J_2$ can be understood qualitatively in the following way.
The ferromagnetism between the nearest Tb is attributed to their short distance (3.54 {\AA}), which is comparable to that of Tb single crystal (3.525 {\AA}), and such short distance leads to the positive exchange integral~\cite{jennings1957heat,thoburn1958magnetic}. 
The ferromagnetism between the next nearest Tb is understood by the superexchange mediated via the Cl-p orbitals. As sketched in Fig. \ref{fig3}(i), the occupied $d_{z^2}$ orbital (orange orbit in the sketch) can be decomposed into $t_{2g}'$ orbitals in the local coordinate as $|d_{z^2}\rangle =\frac{1}{\sqrt{3}}|d_{x'y'}\rangle+\frac{1}{\sqrt{3}}|d_{x'z'}\rangle+\frac{1}{\sqrt{3}}|d_{y'z'}\rangle$. 
Then one can analyze the superechange across the $\backsim$90$^\circ$ Tb-Cl-Tb bonds in $x'y'$, $y'z'$ and $z'x'$ plane respectively.
Without lose the generality, we take the Tb$_1$-Cl-Tb$_2$ in the $x'y'$ plane as an example, as labeled in Fig. \ref{fig3}(i) and also in Fig. S4 of SM~\cite{Zhongsupplemental}. The virtual hopping between 
$p_{x’}$ ($p_{z’}$) of Cl and $d_{x’y’}$ ($d_{y’z’}$) of Tb$_1$ is allowed, which occurs in such a way that the spins on Tb$_1$ are mutually antiparallel. Meanwhile, the remaining $p_{x’}$ ($p_{z’}$) electron of Cl is orthogonal to all the $t’_{2g}$ ($d_{y’z’/x’y’}$) orbitals on Tb$_2$, which lead to the FM exchange integral. Therefore, according the Goodenough–Kanamori–Anderson (GKA) rules~\cite{goodenough1955theory,kanamori1959superexchange,anderson1959new}, the superexchange coupling between Tb$_1$ and Tb$_2$ is dominated by a FM interaction. The similar analysis is applied to $y'z'$ and $z'x'$ plane as shown in Fig. S5 of SM~\cite{Zhongsupplemental}, which present the same results yielding to the $C_{3z}$ symmetry, and finally give rise to a sizable FM $J_2$ comparing to $J_1$.

Additionally, we would like to highlight the significant anisotropy energy $D$=0.31 meV in monolayer TbCl, which is considerably higher than that of 3$d$ compound CrI$_3$ (about 0.2 meV)~\cite{huang2017layer,kim2019giant}. This difference is attributed to the stronger SOC in 5$d$ orbitals compared to 3$d$ orbitals.
The large anisotropy energy can suppress the in-plane magnetic fluctuation, and guarantees the robustly stable out-of-plane magnetic order in the monolayer TbCl. As a result, the sizable FM exchange coupling $J_1$, $J_2$ and anisotropy energy $D$ strongly suggest that the monolayer TbCl is an ideal room temperature FM material.

\begin{figure}[t]
	\centering
	\includegraphics[width=0.48\textwidth]{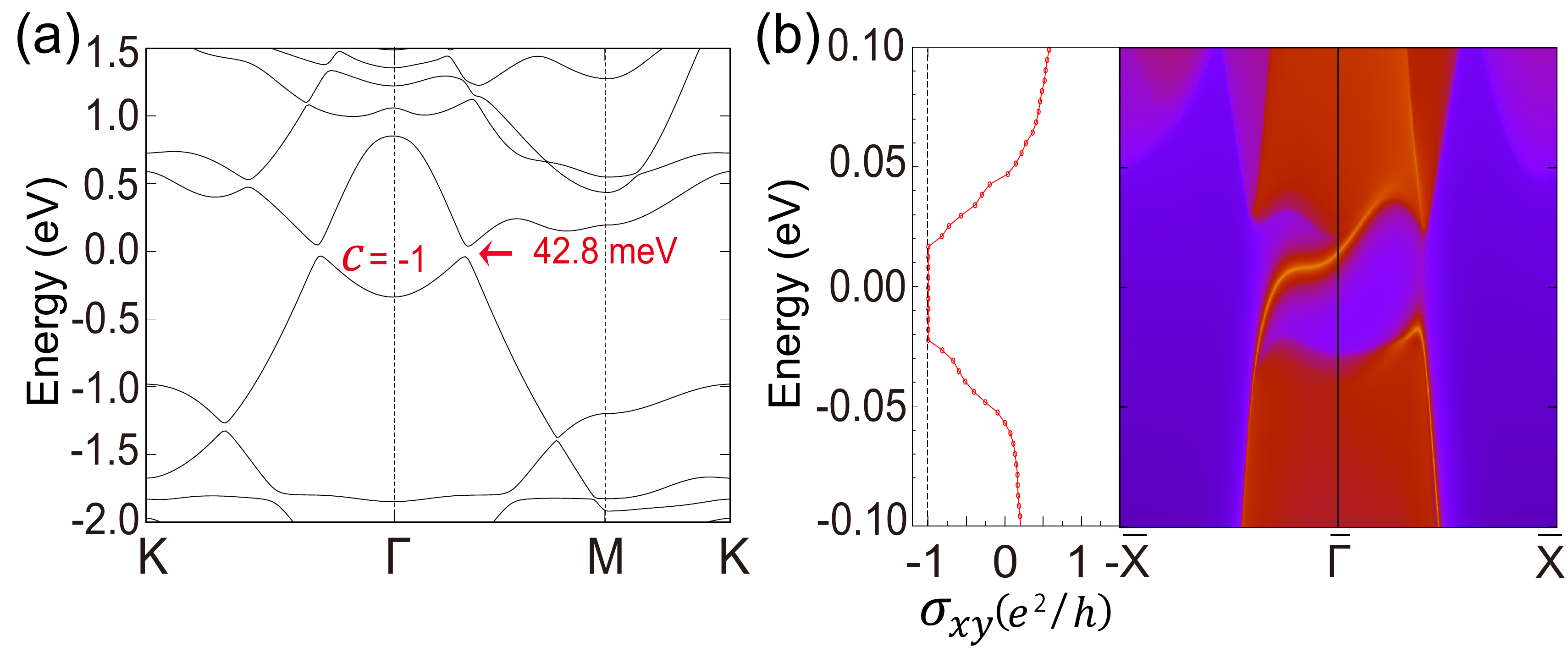}
	\caption{The topological properties of the TbCl monolayer. (a) The band structure of monolayer TbCl with SOC. The band gap and the Chern number are marked. (b) The  left panel is the anomalous Hall conductivity $\sigma_{x y}$ as a function of Fermi energy. The right panel is the chiral edge state on (110) edge. 
	}
	\label{fig4}
\end{figure}

Finally, we determine the band topology and QAH effect of the monolayer TbCl. Based on HSE06 functional method with SOC, the band structure of FM-z monolayer is calculated and demonstrates a large global band gap $\backsim$42.8 meV, as shown in Fig. \ref{fig4}(a). Such large gap is obviously above the room temperature, and is characterized by the Chern number $C=-1$ which is obtained by integrating the Berry curvature of all the occupied bands~\cite{yao2004first,jungwirth2002anomalous}. 
The anomalous Hall conductivity as function of chemical potential is calculated in the left panel of Fig. \ref{fig4}(b), in which a quantized plateau with $\sigma_{x y}=-\frac{e^2}{h}$ appears in the topological band gap, corresponding to the chiral edge state as shown in the right panel.
Thus, based on the significant band gap and high Curie temperature mentioned earlier, the room temperature QAH effect can be achieved in a monolayer of TbCl.

\section{Discussion}
In summary, we predict the monolayer TbCl as the experimental candidate to realize the room temperature QAH effect. The bulk TbCl is found to be a FM insulator supporting the Hall conductivity $G=-e^2/h$ per layer of the crystal. The weak van der Waals interactions allow TbCl to be easily exfoliated into a monolayer, which retains the ferromagnetic properties of the bulk material. And also, the quantum anomalous Hall effect is inherited in the monolayer, corresponding to the Chern number $C=-1$. We estimate the Curie temperature as $\sim$457 K, and the topological gap as $\sim$42.8 meV, which are both beyond the room temperature. Therefore, the monolayer TbCl has the great potential to be a platform to observe and apply the topological quantum phenomena in room temperature. Comparing to the previous theoretically proposed high temperature QAH insulators, monolayer TbCl presents clear advantages. Unlike the most proposals which are based on the 3d orbitals, TbCl is featured by the extended 5d orbitals which induce strong ferromagnetism and a high Curie temperature. Furthermore, its vdW layered structure and low exfoliation energy (0.24 J/m$^2$) make it experimentally accessible, similar to CrI$_3$ and FePS$_3$.


\section{Methods} 
\subsection{Methods for first-principles calculations}
The first-principles calculations are carried out based on density functional theory (DFT), which is implemented in the Vienna ab initio Simulation Package (VASP)~\cite{PhysRevB.50.17953,PhysRevB.54.11169,kresse1999ultrasoft}.
The plane wave energy cutoff is set as 400 eV, and a Monkhorst-Pack k-point mesh of $10 \times 10 \times 10 $ and $20 \times 20 \times 1$ for bulk and monolayer unitcell is used, respectively. 
To accurately capture the interlayer vdW interactions in layered TbCl, we adopted the DFT-D3 scheme, which yields lattice constants and interatomic distances closest to experimental values among the tested vdW correction methods. 
The vacuum layer used for monolayer TbCl is 30 Å. For the correlation effects of the Tb-4$f$ electrons, the Heyd–Scuseria–Ernzerhof (HSE06) hybrid functional and DFT+U approach is adopted in this work~\cite{dudarev1998electron,heyd2003hybrid,krukau2006influence}. 
The calculations of edge state and anomalous Hall conductance are carried out using maximally localized Wannier functions (MLWF) through the Wannier90~\cite{mostofi2008wannier90} and WannierTools packages~\cite{wu2018wanniertools}.
The phonon dispersion spectrum is calculated using a 4 $\times$ 4 $\times$ 1 supercell for monolayer and PHONOPY code with the finite displacement~\cite{togo2015first}. The Curie temperature of monolayer is estimated by Monte Carlo (MC) simulations MCsolver code~\cite{liu2019magnetic}.

\section*{Data availability} All data generated or analyzed during this study are included in thispublished article (and its Supplementary Material).

\section*{Acknowledgments}
This work is supported by the National Key Research and Development Program of China (2024YFA1611200), and the National Natural Science Foundation of China (Grant No. 12274154, 12404182). The computation is completed in the HPC Platform of Huazhong University of Science and Technology. J. Y. Zou also thank the support from Quantum Science Center of Guangdong-Hong Kong-Macao Greater Bay Area GDZX2301005.

\section*{Author contributions} G.X. conceived the project. J.Q.Z. and J.Z.Z. performed the calculations. J.Q.Z., J.Y.Z. and G.X. performed the theoretical analysis. All authors contributed to the writing of the manuscript.

\section*{Competing interests} The authors declare no competing interests. 


%

\newpage

\begin{table} [t]
	\centering
	\caption{The magnetic space group, spin magnetic moment and total energies of bulk TbCl in different magnetic configurations. $U=6$ eV is taken in the DFT+U calculation. The ground state energy is taken as the reference (zero point). }
	\label{tab1}
	\renewcommand\arraystretch{1.4}
	\setlength{\tabcolsep}{0pt}
	\vspace{1.2em}
	\begin{tabular}{l@{\hspace{2mm}}cp{2.7cm}<{\centering}@{\hspace{1mm}}cp{1.8cm}<{\centering}@{\hspace{1.5mm}}cp{1.5cm}<{\centering}@{\hspace{2mm}}cp{1.5cm}<{\centering}@{\hspace{2mm}}cp{1.2cm}<{\centering}@{\hspace{2mm}}cp{1.8cm}<{\centering}@{\hspace{1mm}}cp{1.2cm}<{\centering}@{\hspace{1mm}}cp{1.5cm}<{\centering}}
		\hline\hline
		\rule{0pt}{5pt}
		Config. & MSG  & Tb$_{1}(\mu_{B})$ & Tb$_{2}(\mu_{B})$ & E(meV/$f.u.$) \\
		\hline
		FM-z & $ R$-$3m^{'} $ & (0,0,6.360) & (0,0,6.360) & 0        \\
		FM-x & $ C2/m $ & (6.356,0,0) & (6.356,0,0) & 4.22        \\
		AFM-z & $R$-$3^{'}m^{'} $ & (0,0,-6.239) & (0,0,6.239) & 58.24        \\
		AFM-x & $ C2^{'}/m $ & (-6.236,0,0) & (6.236,0,0) & 59.57        \\
		\hline\hline
	\end{tabular}
	\renewcommand\arraystretch{1.4}
	\label{table:mag}
\end{table}

\begin{table} [t]
	\centering
	\caption{The relative total energy of four different magnetic structure (meV per unit cell), and the estimated exchange coupling $J_1$, $J_2$ (meV), anisotropy energy $D$ (meV/Tb) and Curie temperature $T_\text{c}$ (K). }
	\label{tab2}
	\renewcommand\arraystretch{1.4}
	\vspace{1.2em}
	\begin{tabular}{{l}cp{1cm}<{\centering}cp{1cm}<{\centering}cp{1.1cm}<{\centering}cp{1.1cm}<{\centering}cp{1.1cm}<{\centering}cp{1.1cm}<{\centering}cp{1.1cm}<{\centering}cp{1.1cm}<{\centering} }
		\hline\hline
		\rule{0pt}{13pt}
		&FM-z  &Neel-z &Zigzag-z &120$^\circ$ &$J_1$  &$J_2$  &$D$  &$T_\text{c}$ \\
		\hline
		& 0  &$370.90$  &$534.10$  &$359.61$    &$2.52$  & $2.09$  &$0.31$ &457\\
		\hline\hline
	\end{tabular}
	\renewcommand\arraystretch{1.4}
	\label{table:mag}
\end{table}

\end{document}